\documentclass[twocolumn,tighten]{aastex63}

\usepackage{times,natbib,graphicx,amsmath,multirow,xspace}
\usepackage{xcolor}
\usepackage{lineno}
\usepackage{appendix}
\usepackage{rotating}
\usepackage{longtable}

\begin{document}
\title{Ne and Fe abundances in the ISM: Archival Study of Fe-L and Ne-K edges in Chandra and XMM-Newton}

\correspondingauthor{D. L. Moutard}
\email{moutard@umich.edu}

\author[0000-0003-1463-8702]{~D.~L.~Moutard}
\author[0000-0002-5466-3817]{~L.~R.~Corrales}
\affiliation{Department of Astronomy, University of Michigan, 1085 S. University, Ann Arbor, MI 48109, USA}
\author[0000-0002-1049-3182]{~I.~Psaradaki}
\affiliation{European Space Agency, European Space Research and Technology Center, Keplerlaan 1, 2201 AZ Noordwijk, Netherlands}
\author{~E.~Temple}
\affiliation{School of Physics \& Astronomy, Rochester Institute of Technology, 1 Lomb Memorial Drive, Rochester, NY 14623}
\author{~M.~Shi}
\affiliation{Computer Science and Engineering Division, University of Michigan, 2260 Hayward Street, Ann Arbor, MI 48109, USA}

\begin{abstract}
    The abundance of elements in the interstellar medium (ISM) is a key facet for many fields of astrophysical study. In the soft X-ray spectra, absorption by interstellar gas can result in deep absorption features that affect continuum measurements. In this paper, we focus on measuring the abundance of interstellar iron and neon from the column densities observed in soft spectra from XMM-Newton and Chandra for various low mass X-ray binaries (LMXBs), which allows for a direct probe of elemental abundances. As a noble gas, neon will not deplete into solid form, thus providing a benchmark with abundances determined via UV spectroscopy. We find that, when assuming Fe is 90\% depleted into grains, [Fe/Ne]$ = -0.523\pm0.025$, [Fe/H]$ + 12 = 7.482\pm0.016$,  and [Ne/H]$ + 12 = 8.012\pm0.022$, which are the tightest observational constraints on these abundances to date, while being consistent with literature which uses protosolar abundances. We also test how depletion into solid grains and scattering affect the results. The choice of depletion fraction can affect the abundance measurement by roughly $5\%$, and that the inclusion of a scattering component can affect abundance measurements by $\sim1-7\%$. 
\end{abstract}

\section{Introduction} \label{sec:intro}
\setcounter{table}{0}
The abundances of various elements in the modern Milky Way interstellar medium (ISM) is a crucial benchmark for understanding nucleosynthesis, stellar and planetary lifecycles, and galactic evolution over cosmic time. Soft X-ray spectra ($< 2$~keV) show photoelectric absorption and scattering resonances that are sensitive to the abundance of all the major elements (C, N, O, Ne, Mg, Si, Fe), in both gas and solid phases. Furthermore, the continuum absorption and scattering effects of the ISM are highly sensitive to the assumed abundance tables \citep{wilms00}. By measuring the depth of photoelectric absorption edges in high resolution, we can determine the absolute column density of a particular element in a given X-ray sightline. In this work, we apply this principle to a wide sample of LMXBs to measure the absolute abundances of iron (Fe) and neon (Ne) in the Milky Way ISM.

The effect of ISM absorption on the soft X-ray spectrum has been reported and studied for many years, and efforts have been made to account for these effects and improve upon earlier models of the ISM \citep{ride77}. Some groups use solar  system abundances as an analog to the broader local ISM, using both meteoritic surveys and spectral studies of the photosphere of the Sun (e.g. \citealt{anders89, grevesse98, wilms00, lodders03, asplund09, lodders09, asplund21}). Solar system abundances are a reasonable stand in for local ISM abundances, assuming our solar system contains relatively ``average" makeup. However, as discussed by \cite{nieva12}, there is clearly some bias toward preferentially treating the nearest system as a reference for galactic ISM which has evolved and been seeded with heavier elements since the formation of our solar system. To address this, other groups have turned to soft  measurements of X-ray binaries (XRBs), which can illuminate the ISM and directly allow measurements of elemental abundances along the line of sight \citep{juett04, pinto13, psaradaki24}, and can even allow for the probing of detailed grain composition \citep{psaradaki23, corrales24}. 

It is typical in these studies to compare the measured abundance of each element to that of hydrogen, due to its simplicity and high abundance. In many existing spectral models, some table of abundances relative to hydrogen must be used to estimate the effect of absorption features (e.g. {\sc TBabs} in {\sc xspec} \citep{arnaud96}, which by default assumes the abundance from \citealt{wilms00}, and {\sc absm} in {\sc spex} which assumes the abundances measured in \citealt{morrison83}). In this paper, we focus on the abundance of Fe in the ISM, using Ne as a benchmark rather than hydrogen. Ne is useful for this purpose because of its status as a noble gas; the low affinity for chemical interaction means that Ne will not deplete into grains at a significant rate \citep{gatuzz15}. We analyze low-mass X-ray binaries (LMXBs), using their bright X-ray emission as a means of illuminating the ISM. Spectroscopic instruments with high resolution in the soft X-ray band (such as XMM-Newton and Chandra) will often display an Fe-L absorption edge at approximately 700 eV and a Ne-K absorption edge at roughly 870 eV \citep{juett04}. With these edges clearly resolved in the spectra of the LMXBs, it is possible to measure column density of these elements from the absorption along that line of sight. 

This paper is organized as follows; In Section \ref{sec:data} we describe the data we use and the reduction process. In Section \ref{sec:methods} we describe the fitting and analysis process, and demonstrate the results of the these abundance measurements. Finally, we conclude and summarize our findings in Section \ref{sec:conc}.

\section{Data Reduction} \label{sec:data}
The data used for this study come from archival observations from XMM-Newton and Chandra HETG. We select from X-ray binary sources which have a relatively high column density (typically above $10^{21}$ cm$^{-2}$), allowing for higher signal in the regions of the Fe and Ne edges. To this effect, we also select the sources with the highest exposures among the available catalogs, and focus on those sources with the highest flux. A full list of sources and observations can be seen in Table \ref{tab:obs}. This table only shows the observations which were used in the final analysis; quality cuts based on fit statistics and uncertainties eliminated certain observations from our analysis. These cuts will be described in more detail in Section \ref{sec:methods}.
\begin{table*}[h!]

\centering

\caption{Observation Information}

\begin{tabular}{l|ccc|cc|c}
Source Name & Instrument & Obs. ID & Exposure (ks) & Model & Energy Range (keV) & $N_{HI}~(10^{21} ~{\rm cm}^{-2})$\\
\hline\hline
\multirow{5}{*}{GX 9+9} & \multirow{3}{*}{XMM} & 0694860301 & 49.4 & $^\dagger$ & 0.45 - 1.23 & \multirow{5}{*}{1.90}\\
& & 0090340601 & 21.4 & $^\dagger$ & $^\dagger$ &\\
& & 0090340101 & 2.0 & $^\dagger$ & $^\dagger$ &\\\cline{2-6}
& \multirow{2}{*}{Chandra} & 703 & 20.7 &diskbb+powerlaw & 0.61 - 1.5 & \\
& & 11072 & 95.8 & $^\dagger$ & 0.61 - 1.3 & \\
\hline
\multirow{5}{*}{Cyg X-2} & \multirow{2}{*}{XMM} & 0303280101 & 63.2 & $^\dagger$ & 0.52 - 1.5 & \multirow{5}{*}{1.90}\\
& & 0111360101 & 32.7 & $^\dagger$ & 0.51 - 1.5 &\\\cline{2-6}
& \multirow{2}{*}{Chandra}& 10881 & 66.4 & $^\dagger$ & 0.62 - 1.3 &\\
& & 1102 & 17.6 & diskbb & 0.62 - 1.35& \\
\hline
\multirow{2}{*}{GX 349+2}& XMM & 0506110101 & 50.6 & $^\dagger$ & 0.45 - 1.15 &\multirow{2}{*}{4.66}\\\cline{2-6}
& Chandra & 3354 & 25.7 & $^\dagger$ & 0.69 - 1.5 &\\
\hline
\multirow{5}{*}{GRO J1655-40 }& \multirow{5}{*}{XMM} & 0112921301 & 84.7 & $^\dagger$ & 0.47 - 1.5 & \multirow{5}{*}{5.08}\\
 & & 0112921401 & 31.0 & $^\dagger$ & $^\dagger$ &\\
& & 0112921501 & 30.9 & $^\dagger$ & $^\dagger$ &\\
& & 0155762501 & 46.3 & $^\dagger$ & $^\dagger$ &\\
& & 0155762601 & 43.6 & $^\dagger$ & 0.45 - 1.1 &\\
\hline
XTE J1118+480 & Chandra & 1701 & 23.5 & $^\dagger$ & $^\dagger$ & 0.13\\
\hline
\multirow{5}{*}{4U 1735-44 }& \multirow{2}{*}{XMM}& 0090340201 & 21.2 & $^\dagger$ & $^\dagger$ & \multirow{5}{*}{2.75}\\
& & 0693490201 & 153.2 & $^\dagger$ & $^\dagger$ &\\\cline{2-6}
& \multirow{3}{*}{Chandra} & 704 & 24.4 & $^\dagger$ & 0.62 - 1.5 &\\
& & 6637 & 24.1 & $^\dagger$ & 0.65 - 1.5 &\\
& & 6638 & 23.0 & $^\dagger$ & 0.66 - 1.5 &\\
\hline
\multirow{12}{*}{GX 339-4 }& \multirow{8}{*}{XMM}& 0654130401 & 69.6 & $^\dagger$ & 0.45 - 1.2 &\multirow{12}{*}{3.89}\\
& & 0605610201 & 66.2 & $^\dagger$ & $^\dagger$ &\\
& & 0204730201 & 265.3 & $^\dagger$ & $^\dagger$ &\\
& & 0204730301 & 268.9 & $^\dagger$ & $^\dagger$ &\\
& & 0156760101 & 142.7 & $^\dagger$ & $^\dagger$ &\\
& & 0148220201 & 40.6 & $^\dagger$ & $^\dagger$ &\\
& & 0692341401 & 43.6 & $^\dagger$ & $^\dagger$ &\\\cline{2-6}
& \multirow{4}{*}{Chandra}& 4569 & 49.9 & $^\dagger$ & 0.69 - 1.5 &\\
& & 4570 & 44.5 & $^\dagger$ & 0.68 - 1.35 &\\
& & 4571 & 43.4 & $^\dagger$ & 0.68 - 1.35 &\\
& & 4420 & 74.1 & $^\dagger$ & 0.62 - 1.3 &\\

\hline
\multirow{4}{*}{Ser X-1 }& \multirow{3}{*}{XMM} & 0084020401 & 4.3 & $^\dagger$ & $^\dagger$ & \multirow{4}{*}{4.40} \\
& & 0084020501 & 3.6 & $^\dagger$ & $^\dagger$ \\
& & 0084020601 & 9.2 & $^\dagger$ & $^\dagger$ \\\cline{2-6}
& Chandra & 700 & 76.4 & $^\dagger$ & 0.62 - 1.5 \\
\hline
\multirow{7}{*}{4U 1636-53 }& \multirow{6}{*}{XMM}& 0500350401 & 76.3 & $^\dagger$ & $^\dagger$ & \multirow{7}{*}{3.01}\\
& & 0764180201 & 78.3 & $^\dagger$ & $^\dagger$ &\\
& & 0764180301 & 59.4 & $^\dagger$ & $^\dagger$ &\\
& & 0764180401 & 71.7 & $^\dagger$ & $^\dagger$ &\\
& & 0606070101 & 15.3 & $^\dagger$ & $^\dagger$ &\\
& & 0606070301 & 83.0 & $^\dagger$ & $^\dagger$ &\\\cline{2-6}
& Chandra & 105 & 29.4 & $^\dagger$ & 0.62 - 1.5 &\\
\hline
IGR J00291+5934 & XMM & 744840201 & 170.6 & $^\dagger$ & $^\dagger$ & 4.13\\
\hline
\end{tabular}
\label{tab:obs}

$^\dagger$ Default model values ({\sc powerlaw} from 0.45-1.5 keV). {\bf All } $N_{HI}$ values listed here are from the HI4PI HI survey map, accessed via HEASARC $N_H$ tool,  and do not represent $N_H$ measured in this study, but are instead used as a point of comparison between sources. 
\end{table*}

The Chandra observations were reduced using the web service {\sc tgcat} \citep{huenemoerder11}. This service allows users to search targets by name, and download a version of the observation using a default reduction pipeline. For almost all Chandra observations, we use the medium energy grating (MEG) +1 and -1 orders, and fit them simultaneously. The exception is XTE J1118$+$480, which only had the low energy grating (LEG) spectra available. The MEG spectra are required since these extend to low enough energies to measure the O and Fe edges. 

XMM observations were reduced using the standard XMM pipeline, with {\sc xmmsas v.21.1.0}. Observations were selected and installed from the XMM-Newton science archive. Index calibration files are generated using {\sc cifbuild}, then {\sc odfingest} is used to create a summary file from the provided ODF file from the science archive. We use {\sc rgsproc} to generate the default FITS files, then run {\sc evselect} with time bins of 100s to determine regions of high background. These regions are filtered using {\sc tabgtigen} with the expression {\sc RATE $<$ 0.3} for all sources, creating good time intervals (GTIs) with low background. We re-run {\sc rgsproc} with the newly generated GTI, then combine the RGS1 and RGS2 spectra using {\sc rgscombine}. The resulting unbinned spectra are used in this analysis, alongside the background files which are automatically generated by {\sc rgsproc}. 

Certain types of sources are excluded from this analysis. Specifically, a handful of ultra-compact X-ray binaries (UCXBs) were initially included, but these sources tend to have additional emission in the region around 0.7 keV, attributed to a reflection feature caused by excess oxygen in the accretion disk \citep{koliopanos13,koliopanos21}. This excess emission contaminates the Fe-L edge, making a confident measure of the abundance impossible without very careful reflection modeling.  

Metallicity in the the Milky Way is not uniform, with higher metallicities found closer to the center, and other inhomogeneities throughout. LMXBs used in this study lie at distances $\lesssim6.5$ kpc. The resulting hydrogen column densities $N_H$ are comparable to the high end of those studied in \cite{jenkins09}, which used UV observations of stars in a more local search. By comparison, this investigation is a general study of broader galactic abundances, though there is some overlap in density.

\section{Abundance Measurements} \label{sec:methods}
Each spectrum is fit in the 0.45-1.5 keV range by default. The lower end of this range allows for the fitting of the O-K edge at roughly 0.543 keV, and provides continuum information in the regions between the O-K and Fe-L edges. The upper end of this range provides additional continuum information to ensure that the photoelectric absorption continuum is accurately measured. All spectral fitting is done using {\sc xspec v.12.14.1} and {\sc v.12.5.0}\footnote{{\sc xspec} was updated mid-analysis, though tests indicate this does not affect the final results}\citep{arnaud96}. The default model is {\sc tbvarabs*ISMabs*xscat*powerlaw}, using the {\sc wilm} abundance \citep{wilms00} and {\sc vern} cross-sections\citep{verner96}. A simple power law is not typically a physically motivated model for the soft X-ray spectrum of X-ray binaries, but works locally to describe a small region of the continuum. 

{\sc TBVarabs} is a model that accounts for interstellar absorption while allowing elemental abundance as a free parameter. It also allows users to vary the depletion factor $(1 - \beta_Z)$ \citep{wilms00}. $\beta_Z$ represents the number of atoms per element $Z$ that are in solid grains, in contrast to those that are in the gas phase, with a value of $(1 - \beta_Z) = 1$ meaning the element is entirely in gas phase, and $(1 - \beta_Z) = 0$ meaning the element is found only as part of solid grains. As elements deplete into grains, their absorption profile is affected by self shielding, which produces shallower absorption edges \citep{wilms00}. By default, \texttt{XSPEC} sets all depletion factors to 1. We adjusted the depletion values for all elements following the values from Table~2 of \cite{wilms00}. This term modulates the gas absorption cross section from \cite{wilms00}
\begin{equation}
    \sigma_{gas} = \sum_{Z,i}A_Za_{Z,i}
    (1 -\beta_{Z,i})\sigma_{\rm bf}(Z,i)
\end{equation}
where $A_Z$ is the abundance of element $Z$ with respect to the number of H atoms in a sightline, $a_Z$ is the fraction of element $Z$ in a particular ionization stage $i$, and $\sigma_{\rm bf}(Z,i)$ is the photoionization cross section of element $Z$ in ionization stage $i$.

{\sc ISMabs} is a model component that provides high resolution absorption templates for neutral and ionized species of many common ISM elements \citep{gatuzz15}. We set the Ne abundance in \texttt{tbvarabs} to zero and replaced it with the \texttt{ISMabs} model. Following the measurements of \citep{gatuzz15}, we only fit for the column density of Ne~{\sc I} and Ne~{\sc II}, which accounts for $> 98\%$ of the interstellar neon. All other elements in the \texttt{ISMabs} model are frozen to 0. When fitting for the Ne components, we limit the passband around the Ne-K edge more directly. XMM observations are fit between 0.8$-$1.2, and Chandra observations are limited to between 0.8$-$1.0 keV, to avoid pileup common above this energy. 

The {\sc xscat} model accounts for scattering by dust \citep{smith16}. This model includes small high resolution features that contribute to absorption edge structure \citep[e.g.,][]{corrales16}. We tie the $N_{\rm H}$ value in {\sc xscat} to that of {\sc TBVarabs}. The circular extraction region \texttt{R\_Ext} is set to 1.5~arcsec for Chandra observations and 10.0~arcsec for XMM-Newton observations, which correspond to the approximate point-spread functions of each instrument. We set the {\sc xpos} parameter, which defines the location of dust scattering, to 0.5. This corresponds to scattering by dust exactly halfway between the observer and the source, which is a reasonable average assumption. This is demonstrated in \cite{corrales16}, which shows that varying {\sc xpos} when using an \texttt{R\_Ext} as small as 10" makes very little difference to the absorption cross section unless using values at the extremes, since very little of the scattered light will be recaptured.

This default model and energy range is used as often as possible for consistency, but it is not appropriate for every source or observation. For example, in some cases a particularly soft spectrum is better described by a disk blackbody component {\sc diskbb}. In cases where the default model is not used, this is indicated in Table \ref{tab:obs} (without the {\sc TBVarabs*xscat} portion labeled to maintain readability). Frequently, the softest bands of Chandra observations are noise dominated, and so those regions are removed.

\begin{figure*}
    \centering
    \includegraphics[width=0.98\linewidth]{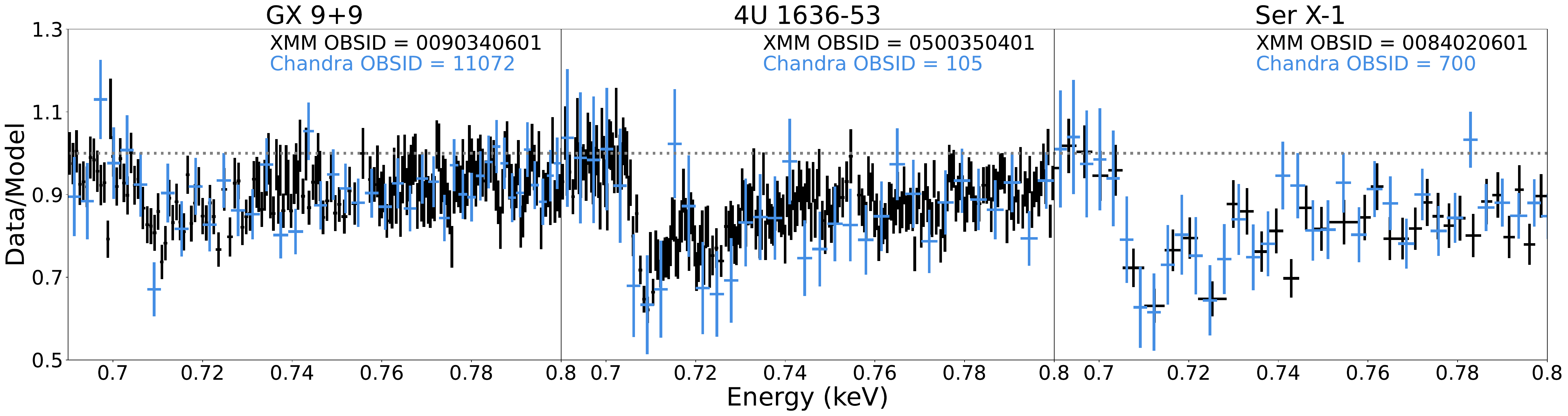}
    \caption{A selection of both XMM and Chandra Fe-L edges with varying column densities. These plots are constructed by fitting the data, then evaluating the model with the Fe abundance set to 0 and plotting the resulting ratio of the data to the model. From left to right, these span a low ($N_H = 1.90\times10^{21}$ cm$^{-2}$), moderate ($N_H = 3.01\times10^{21}$ cm$^{-2}$), and high  ($N_H = 4.40\times10^{21}$ cm$^{-2}$) column density,  as reported by the HEASARC HI4PI tool. this demonstrates how the depth of the edges vary as the column density increases.} 
    \label{fig:repspec}
\end{figure*}

The observations shown in Table \ref{tab:obs} are those observations which passed a quality cut after fitting the spectra. This quality cut eliminated any spectrum which had a reduced fit statistic (C-stat per degree of freedom) \footnote{While the fit statistic reported by {\sc xspec} is called C-stat, when the background spectra generated by {\sc rgsproc } are loaded, the statistic used is actually the related W-stat. For more information, see https://heasarc.gsfc.nasa.gov/docs/software/xspec/manual/node340.html} of three or greater for this initial continuum fit (that is, the fit with all abundances initially fixed to 1). These cuts also eliminated any observations whose abundance measurements had negative errors (an indication that the fits in {\sc xspec} were not sensitive to the parameter, and thus could not be confidently extracted). We display a handful of representative spectra and their continuum fits in Figure \ref{fig:repspec}. From left to right these represent a source with increasing $N_H$, encompassing a low, moderate, and high column density. The $N_H$ values shown here are from Table \ref{tab:obs}, which we expect to be lower than measured $N_H$ values. We choose these values in comparisons of sources since they represent a uniform $N_H$ measurement and the relative differences between source columns are still apparent. We see that the edges become more pronounced as $N_H$ increases. For a detailed view of the fit parameters and retrieved abundances, see the Appendix.

\subsection{Abundances from Fit} \label{subsec:abunds}


After the initial continuum model fit, we let the Fe abundance parameter be free. The column density of Fe was measured from the resulting best fit parameters, following
\begin{equation}
    N_{\rm Fe} = C_{w}A_{{\rm Fe}}N_{\rm H}
\end{equation}
where $A_{\rm Fe}$ is a free parameter and $C_w$ is the relevant scale factor from the{\sc wilm} abundance table in {\sc xspec}. ($C_w = 2.69$ provides the column density of Fe  in units of $10^{17}$ cm$^{-2}$). 
We measured $N_{\rm Fe}$ with different assumed depletion values, including 1 (no depletion), 0.1 (10\% depletion) and 0 (100\% depletion). The last two choices are motivated by UV depletion studies \citep{jenkins09} and other X-ray measurements \citep{psaradaki23, psaradaki24}. Note that  \texttt{ISMabs} uses the cross-section for metallic iron from \citet{kortright00}, and thus does not provide accurate modeling for atomic Fe. Changing the depletion factor in this model only changes the optical depth to self-absorption.


The Ne column is returned directly by {\sc ISMabs}. Due to its high flexibility and independence, we found that the {\sc ISMabs} model is more sensitive to continuum variations than {\sc TBvarabs}. The reported Ne column is the sum of the Ne {\sc I} and Ne {\sc II} values. Throughout this work, we assume that the Fe and Ne absorption features, on average, arise from the same sources within the diffuse ISM. There is no evidence for cold or near-neutral absorbers within the high energy environment of LMXBs. Therefore, we take the ratio of the elemental column densities found in the X-ray spectra as a measurement for the relative abundances of various elements in the ISM. After the final fitting for Fe and Ne column densities, we also noted the final H column reported by \texttt{tbvarabs} as a point of reference. We also test the validity of excluding Ne III by re-calculating the abundances with the assumption that our measurements are underestimated. Between \cite{gatuzz15} and \cite{gatuzz16}, the sightline with the greatest contribution of Ne III is roughly 5.5\%. We take the extreme case and assume all of our abundances are underestimated by this amount. We find that the [Fe/Ne] and [Ne/H] measurements still agree within uncertainty.

\begin{figure*}[t!h!]
    \centering

    \includegraphics[width=\linewidth]{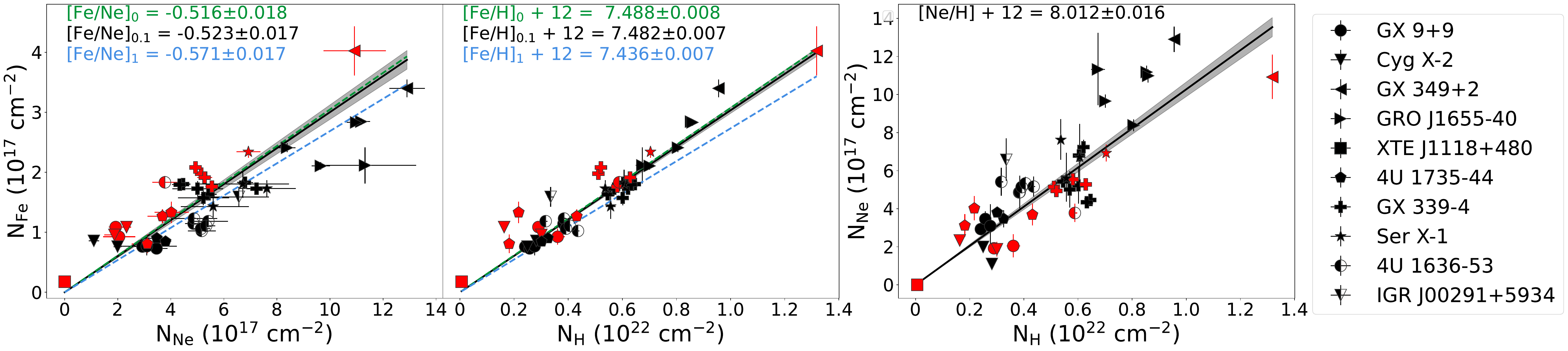}

    \caption{The abundance of Fe compared to Ne (left) and H (center). We also compare Ne to H (right). Black points represent XMM-Newton observations, and red points represent Chandra observations. The data points shown on this graph are only for the fiducial Fe depletion of 0.1. The fit to that data set is shown as a solid black line, with the gray shaded region being the uncertainty. The green and blue dashed lines represent fits to data sets with Fe depletion of 0 and 1, respectively. We do not plot the data sets for these depletion factors, nor do we plot the uncertainties in their fit to maintain readability. The uncertainties reported in this plot are only those from the fit, and do not include additional uncertainty from statistical comparison of models.m The low column source XTE J1118+480 provides only an upper limit on ${\rm N_{Ne}}$}.
    \label{fig:abunds}
\end{figure*}

We plot the measured Fe column ($N_{\rm Fe})$ against both $N_{\rm Ne}$ and $N_{{\rm Fe}}$ in Figure~\ref{fig:abunds}. As the Ne and H column densities increase, the Fe abundance also increases. We are then able to measure the relative abundances of each element by measuring the slope of the data points. We fit each combination of column densities with a linear model and a y-intercept fixed at 0. This fit is done using the orthogonal distance regression (ODR) method in {\sc scipy v. 1.13.1}, which accounts for uncertainty along both axes. The results of these fits can be seen in the black lines in Figure \ref{fig:abunds}. The error bars on the data represent a 1-$\sigma$ uncertainty, and the gray region represents the 1-$\sigma$ deviation in the ODR fit. 

As can be seen in Figure~\ref{fig:abunds} and the tables in the appendix, there is occasionally some variation in fit parameters for the same source. The largest of these variations are seen between instruments, which is largely driven by the softest bands of the Chandra data being lower signal than the same bands in XMM. This can affect the precision of the $N_H$ measurements, and the overall continuum shape. Similarly, the detailed spectral state, which our bandpass is not broad enough to tightly constrain, can also drive some of these variations. Regardless, the final measured columns typically agree within uncertainty. 

\begin{figure}[t!h]
    \centering
    \includegraphics[width=\linewidth]{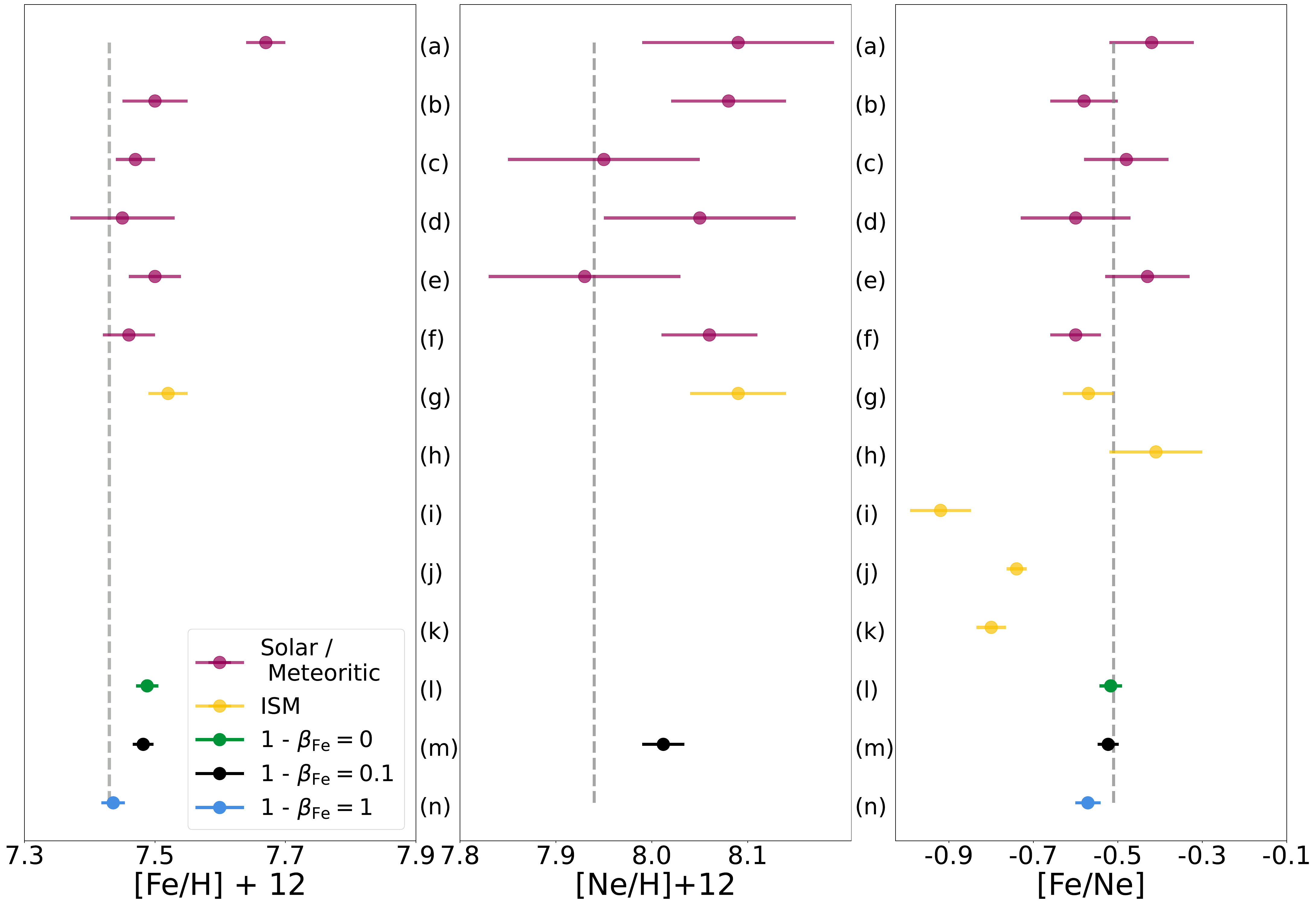}
    \caption{Here we compare our fit results to those measured in other literature. In order, these points refer to (a) \cite{anders89}, (b) \cite{grevesse98}, (c) \cite{lodders03}, (d) \cite{lodders09}, (e) \cite{asplund09}, (f)\cite{asplund21} , (g) \cite{nieva12}, (h) \cite{psaradaki24}, (i) \cite{gatuzz15}, (j) \cite{gatuzz16}, and (k) \cite{juett06} Rows (l), (m), and (n) are the results of this work for  $1 - \beta_{{\rm Fe}}=0,0.1,$ and 1, respectively. The vertical line is the abundance put forward by \cite{wilms00}. The error bars displayed here represent a conservative estimate on the uncertainty, which includes the discrepancy between models with and without scattering.}
    \label{fig:litcomp}
\end{figure}

We compute the relative abundances using standard notation, where brackets indicate the logarithm of the ratio of the abundances plus 12. 
We find a best-fit value of [Fe/Ne] = $-0.523\pm0.017$,  [Fe/H] + 12 = $7.482\pm0.007$ and [Ne/H] + 12 = $8.012\pm0.016$. These values are consistent with those measured by previous studies (Figure~\ref{fig:litcomp}), but with significantly smaller error bars. 

\subsection{Fit Result Validation} \label{subsec:validation}

To test the validity of our error bars, we apply a jackknife resampling method, removing one data point at a time and re-fitting for the parameterized abundance. We find that resampling yields the same slope on average, with a standard deviation on the slope that is a factor of 4 smaller than the uncertainty provided by the ODR method. This leads us to conclude that the fits are not largely skewed by any individual points, and that the 1-$\sigma$ uncertainties returned by ODR are acceptable. 

To further validate these uncertainties we expand upon the jackknife resampling method. We randomize the order of the data and remove three points at a time rather than one, and re-fit for the relative abundances. We repeat this process with the entire data set 20 times, and find similar results to the single-point jackknife resampling; the median values still match the initial fit value with a standard deviation 3-4 times smaller than the uncertainty in the fit. 

We also repeated the entire process described in Section \ref{subsec:abunds}, but using the {\sc TBvarabs} model to measure the Ne column density (by letting $A_{\rm Ne}$ be a free parameter). For neon, $C_w = 8.71$. We find that the measured Ne column in {\sc TBvarabs} is systematically higher by $\sim2\%$, but the resulting abundance fits agree closely to our \texttt{ISMabs} results within the uncertainty. This provides an additional validation of our results, indicating that these measured absolute abundances are consistent across models.

\subsection{Scattering and Shielding} \label{subsec:scatshield}

Figures~\ref{fig:abunds} and \ref{fig:litcomp} also show how both depletion into grains and scattering affect the measured abundance ratios. We take $(1 - \beta_{\rm Fe}) = 0.1$ as our fiducial model. 
The blue and green dashed lines in Figure \ref{fig:abunds} represent a fit to abundance measurements with the \texttt{tbvarabs} depletion parameters set to 1 and 0, respectively. We do not display the individual data points in this plot to avoid cluttering the fiducial data, but we still include the fit value and uncertainty in matching text color in the plot, and labeled with a subscript matching the value of $\beta_{{\rm Fe}}$. 

\begin{figure*}[t!h]
    \centering
    \includegraphics[width=0.98\linewidth, trim = 0 0 0 0, clip]{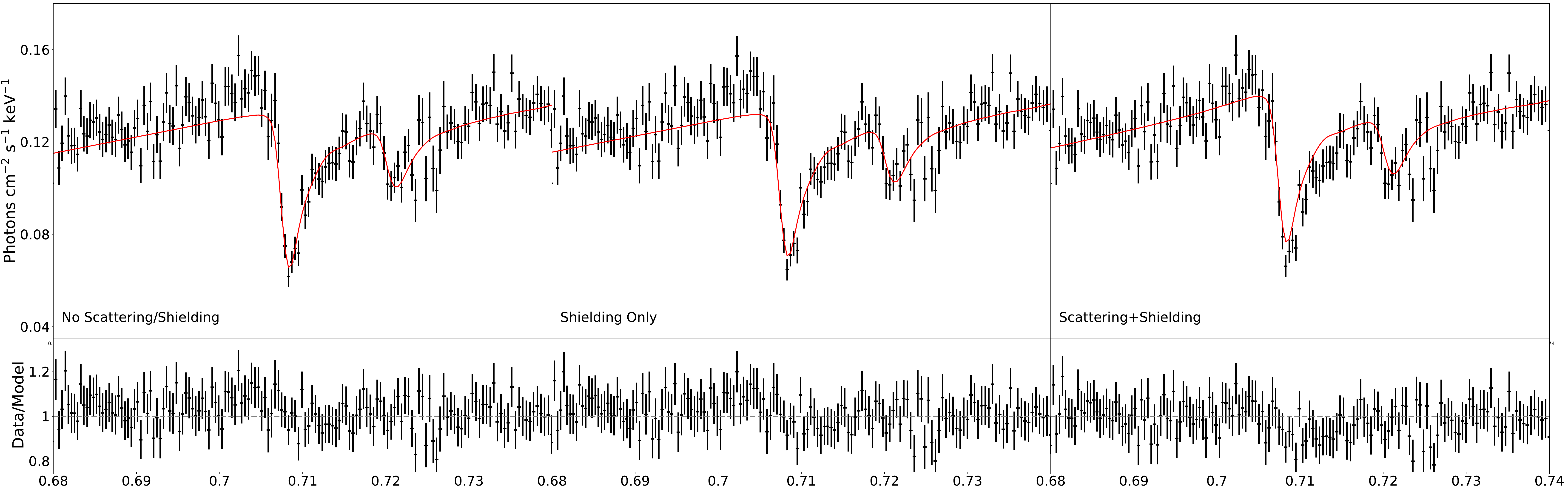}
    \caption{(Top) The 0.68-0.74 keV region surrounding the Fe-L edge for 4U 1636-53 (XMM Observation 0500350401). (Bottom) The data divided by the fit model, which demonstrate the slight variations between the various models. The leftmost panel demonstrates the fit with no scattering or shielding due to depletion. The center panel is identical, but the depletion fractions are set to physically realistic values. The right panel shows the fit with both scattering and shielding due to depletion. Visually, it appears that scattering improves the fit on the low energy side of the edge, and shielding improves the fit surrounding some of the structures on the high energy end of the fit. This is perhaps more apparent in the residuals along the bottom column, which demonstrates the pre-edge region improving with a scattering component, followed by an underestimate of the edge depth.} 
    \label{fig:scatshield}
\end{figure*}

There is a systematic linear increase in the measured Fe abundance with higher depletion. This makes sense, because self-shielding by dust grains leads to shallower photoelectric absorption for a given column density. We find that when it is assumed that Fe is completely in the gas phase, the fit yields [Fe/Ne]$_1$ = $-0.571\pm0.017$ and [Fe/H]$_1$ + 12 = $7.436\pm0.007$. When we move to the other extreme and assume complete depletion into grains ($\beta_{{\rm Fe}} = 0$), we find a fit value [Fe/Ne]$_0$ = $-0.516\pm0.018$ and [Fe/H]$_0$ + 12 = $7.488\pm0.008$. We apply a linear model of both [Fe/Ne] and [Fe/H] + 12 as a function of $\beta$ and find slopes of $-0.056\pm0.010$ and $-0.050\pm0.003$, respectively. These slopes indicate that the range of possible $\beta$ can yield an approximately $5\%$ difference of measured abundance. While this difference is small, it is larger than the measured uncertainties in the slope, implying that it is a meaningful parameter and should be considered in abundance studies. 

We also test how the scattering component of dust extinction affects our abundance measurements. Figure~\ref{fig:scatshield} shows how including scattering provides an improved fit in the regions immediately before the edge, but does cause a sight mismatch in the region from 0.71-0.72 keV. The inclusion of a scattering component typically causes the fit statistic to worsen. This is unsurprising, as the fits are largely driven by the continuum, and the fine structure is more sensitive to the precise material used to generate the absorption cross section \citep{westphal19, psaradaki23, corrales24}. In this study, we only use a generic Fe cross section. The effect is more significant in XMM observations, which can be attributed to the continuum-driven nature of these fits as well, since Chandra data typically have lower signal-to-noise, yielding less sensitivity to changes in the fine structure. Despite this, we opt to keep the scattering component in the model, as it is more physically accurate. We explore this further by testing the other dust models available in {\sc xscat} as a model parameter. For this study we use the default MRN dust grain distribution \citep{mathis77}, but we also test the WD \citep{weingartner01} and ZDABAS \citep{zubko04} distributions. We find that these changes typically only make a marginal change in the fit statistic, and usually in the direction of a worsening fit. Improvements in the fit statistic are only ever seen using the WD distribution, but these are usually small improvements, so for consistency we opt to continue with the default MRN distribution. 

The inclusion of a scattering component systematically decreases the Fe abundance, while increasing the Ne abundance (Figure~\ref{fig:abunds}). When performing the abundance measurements without scattering, we find that the measured slope for Fe/Ne is the one that is most affected, with a $\sim5-7\%$ discrepancy between models, as calculated by measuring the difference in slope and dividing by the average. The Fe/H and Ne/H slopes are less affected, seeing typically $\sim1-4\%$ discrepancies. 

The fine structure components of Fe L photoabsorption features are sensitive to the chemical composition of the dust grains. A different solid phase cross-section might provide a better fit to the absorption and scattering features \citep{lee09, psaradaki23, corrales24}. Nonetheless, the depth of the photoelectric edge and the absorption continuum will remain the same per Fe atom in the compound \citep[e.g.,][]{corrales25}. Thus we expect different assumptions about the chemical composition of the Fe-bearing dust grains to inject no more systematic uncertainty than is already found here, where we have tested the effect of different scattering and depletion choices.

\begin{table}[]

\centering

\caption{Abundances}
\hspace{-1.5cm}
\begin{tabular}{c|c|cc|cc}
& ($1-\beta_{{\rm Fe}}$)& Value & $\sigma_{tot}$ & $\sigma_{fit}$ & $\sigma_d$\\
\hline\hline
\multirow{3}{*}{[Fe/Ne]} &0 &-0.517 &0.027 &0.018 &0.009\\
&{\bf 0.1} &{\bf -0.523} &{\bf 0.025} & {\bf 0.017} &{\bf 0.008}\\
&1 &-0.571 &0.030 &0.017 &0.013\\
\hline
\multirow{3}{*}{[Fe/H]+12} &0 &7.487 &0.017  &0.008 &0.009\\
&{\bf 0.1} &{\bf 7.481} &{\bf 0.016} &{\bf 0.007} &{\bf 0.008}\\
&1 &7.436 &0.018 &0.007 &0.011\\
\hline
\multirow{3}{*}{[Ne/H]+12} &0 & --- &--- &--- &--- \\
&{\bf 0.1} &{\bf 8.012} &{\bf 0.022} &{\bf 0.016} &{\bf 0.006} \\
&1 & --- &--- &--- &--- \\

\end{tabular}
\label{tab:abunds}

Bold values indicate the fiducial values, which we take as the final abundance measurements.

\end{table}

\subsection{Final uncertainty calculation}

We estimate the total uncertainty of the fit $\sigma_{tot}$ using
\begin{equation}\label{eq:unc}
    \sigma_{tot} = \sigma_{fit} + S_{1}\frac{2|S_{1}-S_{2}|}{S_1+S_2}
\end{equation}
where $\sigma_{fit}$ is the $1\sigma$ error in the fit to the slope, and $S_1$ and $S_2$ are the measured slopes with and without a scattering component, respectively. We will refer to this second term as $\sigma_d$, or the discrepancy between models. The final uncertainty is simply the sum of these terms, to ensure that the errors reported cover the values retrieved when scattering is omitted. The uncertainties and error bars displayed in Figure \ref{fig:litcomp} reflect this conservative uncertainty measurement. Our final abundance measurements with the fiducial value of $(1 - \beta_{\rm{Fe}})=0.1$ is [Fe/Ne]=$-0.523\pm0.025$, [Fe/H]~+~12=$7.82\pm0.016$, and [Ne/H]~+12~=$8.012\pm0.022$. A full table of abundances can be seen in Table \ref{tab:abunds}. Even with these conservative uncertainties, our measurements provide the strongest constraint on these abundances when compared to the literature , as shown in Figure \ref{fig:litcomp}. This robust measurement can be attributed to the number of observations used, as well as the broader continuum range when compared to similar studies.

The values in Figure \ref{fig:litcomp} arise from column densities measured through various means. Rows (a)-(g) all come directly from Table 5 in \cite{psaradaki24}, which is itself row (h). Rows (a) - (f)  all arise from measurements of either proto-solar or meteoritic abundances \citep{anders89,grevesse98, lodders03, lodders09, asplund09, asplund21}. Rows (g) - (k) represent direct measurements of the ISM. Row (g) uses optical measurements of B stars \citep{nieva12}, while (h) uses  FUV and X-ray measurements of Cyg X-2 \citep{psaradaki24}. 

Rows (i) - (k) utilize a multiple observations of X-ray binaries and model the abundances using {\sc ISMabs}, in a similar manner to this work, with some overlap in observations. Generally speaking, we find that our Ne column densities generally agree quite well, though we tend to estimate slightly higher values of $N_{\rm Fe}$, yielding distinct measurements of [Fe/Ne]. This can be attributed to the fact that {\sc ISMabs} does not account for depletion and subsequent self-shielding. This explanation for values lower than solar is also suggested by \cite{juett06} and \cite{gatuzz15}, and is consistent with our findings that the assumption with no depletion yield the lowest values of [Fe/Ne]. There is, however, still disagreement between the Fe column we  measured at $1 - \beta_{{\rm Fe}} = 1$ and those seen in rows (i) - (k) of Figure \ref{fig:litcomp}. We attribute this to the fact that we fit over a larger continuum than these studies, which can provide more significant statistics and constraints on the abundance measurements. 

The measurement of the [Fe/Ne] abundance is the most robust of three. That is because there are no spectral features of neutral H in the X-ray. With X-ray spectroscopy, the H column density is inferred based on the relative mixture of other elements. This inference is particularly sensitive to the relative abundance of C and O, which are the most abundant metals in the ISM in addition to being strong soft X-ray absorbers. Due to their strong extinction power, spectroscopic features from C K (0.3~keV) and O K (0.5~keV) photoelectric edges are more rare to find. Without these observable features, $N_{\rm H}$ measurements via X-ray spectroscopy are subject to considerable systematic uncertainty. Due to the relatively low $N_{\rm H}$ required to retain visible Fe L shell features, the majority of our LMXB spectra have visible O K features that are included in the fit. The measured $N_{\rm H}$ column in these cases are thereby subject to the systematic uncertainty in [O/H]. We reserve study of the oxygen column densities and abundances for a future work.

The uncertainties on the [Ne/H] abundances from the literature, which might be used to convert [Fe/Ne] to [Fe/H], are too large to yield meaningful results when compared to our [Fe/H] measurements via X-ray spectroscopy. Despite being sensitive to systematic uncertainties in the absolute abundances of carbon and oxygen, the $N_{\rm H}$ in our fits by nature of the \texttt{tbvarabs} model are constrained by all metals, not just oxygen. Thus it is not surprising that the [Fe/H] and [Ne/H] values are consistent with the literature. Future studies may require a systematic shift in these final values.

\section{Conclusions}\label{sec:conc}

X-ray spectroscopy provides a means to measure both the gas and solid phase column density of all abundant metals. This is achieved by examining high resolution soft X-ray spectra for various LMXBs. We utilize Fe and Ne photoelectric absorption edges, as well as the continuum absorption, to explore the Fe abundance in the ISM in comparison to both Ne and H. By comparing the measured column densities of each, we can measure the relative abundances among all three. We focus on Ne as a benchmark because of its propensity to remain in the gas phase.  This makes the [Fe/Ne] measurement the one that is least subject to systematic uncertainty. We find [Fe/Ne]=$-0.523\pm0.025$, [Fe/H]~+~12=$7.482\pm0.016$, and [Ne/H]~+12~=$8.012\pm0.022$, providing some of the highest precision measurements to date which account for both depletion into grains and scattering. These discrepancies can likely be attributed to differences in model selection, though a full comparison of models is beyond the scope of this work.  .

We also studied how scattering and self-shielding affect the Fe abundance determination via X-ray spectroscopy. We find that the abundance is in fact sensitive to the choice of depletion fraction. Our fiducial model assumes that 90\% of the interstellar Fe is depleted into dust grains. We note that there is a systematic increase of the measured absolute Fe abundance as the gas fraction varies from 1 (entirely gas phase) to 0 (entirely solid phase) on the order of $5\%$. Scattering can also affect the abundance measurement, systematically increasing the Ne abundance measurement while decreasing the Fe abundance measurement. This discrepancy is taken into account for our final conservative error estimates, ensuring that our error bars cover both values.

{\it Acknowledgements:} This research has made use of data and/or software provided by the High Energy Astrophysics Science Archive Research Center (HEASARC), which is a service of the Astrophysics Science Division at NASA/GSFC. Support for this work was provided by the National Aeronautics and Space Administration (NASA) through Chandra Award Number GO3-24129X, NASA Grants 80NSSC20K0883, 80NSSC18K0978, and 80NSSC25K7064. This paper employs a list of Chandra datasets, obtained by the Chandra X-ray Observatory, contained in the Chandra Data Collection~\dataset[DOI: 10.25574/cdc.520]{https://doi.org/10.25574/cdc.520}

\bibliographystyle{aasjournal}
\bibliography{bibliography}

\appendix
\setcounter{table}{0}
\renewcommand{\thetable}{A\arabic{table}}

\section{Appendix}\label{appendix}

This appendix outlines detailed fit parameters, with Table \ref{tab:contfit} showing the continuum fit values for each observation and Table \ref{tab:abundfit} showing the fit vlaues for Ne, Fe, and $N_H$ from the various absorption models. In Table \ref{tab:contfit}, the normalization of the {\sc diskbb} component is defined as $(R_{in}/$km)$^2/(D/10$ kpc)$^2\times\cos{\theta}$, while the normalization of {\sc powerlaw} is photons keV$^{-1}$ cm$^{-2}$ s$^{-1}$ at 1 keV. The C-stat (dof) column displays the fit statistic and degrees of freedom (dof) for each observation. In Table \ref{tab:abundfit}, Ne{\sc i} and Ne{\sc ii} are thevalues from {\sc ismabs}, while Fe and $N_H$ are the values from {\sc TBvarabs}, with Fe being the dimensionless scale factor from fitting. All values shown here assume a depletion parameter of 0.1 and scattering included.

\begin{table*}

\centering

\caption{Continuum Information}

\begin{tabular}{l|cc|cc|cc|c}
Source Name & Instrument & Obs. ID & \multicolumn{2}{c|}{{\sc powerlaw}} &\multicolumn{2}{c|}{{\sc diskbb}} &  C-stat (dof) \\

& & & $\Gamma$ & norm & kT (keV) & norm & \\
\hline\hline
\multirow{5}{*}{GX 9+9} & XMM & 0694860301 & 1.181$\pm0.028$ & 0.773$\pm0.007$ &  -- & -- & 2412.94 (1707)\\
 & XMM & 0090340601 & 1.346$\pm0.022$ & 0.881$^{+0.009}_{-0.008}$ &  -- & -- & 2238.44 (1882)\\
 & XMM & 0090340101 & 1.422$\pm0.074$ & 0.915$^{+0.03}_{-0.029}$ &  -- & -- & 2145.7 (1882)\\\cline{2-8}
 & Chandra & 703 & 5.941$^{+0.612}_{-0.513}$ & 0.082$^{+0.045}_{-0.034}$ & 0.545$\pm0.016$ & 1840.85$^{+275.72}_{-238.34}$ & 5314.03 (4819)\\
 & Chandra & 11072 & 1.502$^{+0.045}_{-0.044}$ & 0.839$\pm0.017$ &  -- & -- & 4812.81 (4311)\\\hline
\multirow{4}{*}{Cyg X-2} & XMM & 303280101 & 1.431$\pm0.011$ & 1.691$^{+0.009}_{-0.008}$ &  -- & -- & 2698.99 (1511)\\
 & XMM & 111360101 & 1.733$\pm0.015$ & 2.14$\pm0.014$ &  -- & -- & 3758.83 (1557)\\\cline{2-8}
 & Chandra & 10881 & 1.172$\pm0.032$ & 1.781$^{+0.027}_{-0.026}$ &  -- & -- & 6538.99 (4179)\\
 & Chandra & 1102 &  -- & -- & 1.172$\pm0.032$ & 35783.621$^{+2423.644}_{-2284.809}$ & 5359.62 (4321)\\\hline
\multirow{2}{*}{GX 349+2} & XMM & 0506110101 & 1.006$^{+0.123}_{-0.12}$ & 2.173$^{+0.11}_{-0.101}$ &  -- & -- & 2294.14 (1633)\\\cline{2-8}
 & Chandra & 3354 & 3.115$\pm0.063$ & 5.634$^{+0.213}_{-0.204}$ &  -- & -- & 4549.49 (3875)\\\hline
\multirow{5}{*}{GRO J1655-40} & XMM & 112921301 & 1.504$^{+0.121}_{-0.12}$ & 0.064$\pm0.004$ &  -- & -- & 1977.33 (1765)\\
 & XMM & 112921401 & 1.697$\pm0.021$ & 8.357$^{+0.095}_{-0.091}$ &  -- & -- & 4581.82 (1884)\\
 & XMM & 0155762601 & 1.749$^{+0.078}_{-0.076}$ & 6.687$^{+0.2}_{-0.187}$ &  -- & -- & 2654.25 (1579)\\
 & XMM & 112921501 & 1.719$^{+0.022}_{-0.021}$ & 8.562$^{+0.098}_{-0.093}$ &  -- & -- & 4631.21 (1882)\\
 & XMM & 0155762501 & 0.726$^{+0.032}_{-0.031}$ & 6.599$^{+0.1}_{-0.097}$ &  -- & -- & 4611.23 (1880)\\\hline
\multirow{1}{*}{XTE J1118+480} & Chandra & 1701 & 1.968$^{+0.018}_{-0.017}$ & 0.246$\pm0.002$ &  -- & -- & 3508.34 (3081)\\\hline
\multirow{5}{*}{4U 1735-44} & XMM & 0090340201 & 1.449$^{+0.027}_{-0.026}$ & 0.817$^{+0.01}_{-0.009}$ &  -- & -- & 2347.17 (1885)\\
 & XMM & 0693490201 & 1.297$\pm0.012$ & 0.561$\pm0.003$ &  -- & -- & 4218.16 (1887)\\\cline{2-8}
 & Chandra & 704 & 2.139$^{+0.043}_{-0.042}$ & 0.96$\pm0.022$ &  -- & -- & 4972.58 (4687)\\
 & Chandra & 6637 & 0.959$\pm0.046$ & 0.587$^{+0.016}_{-0.015}$ &  -- & -- & 4585.21 (4317)\\
 & Chandra & 6638 & 1.052$\pm0.049$ & 0.634$^{+0.018}_{-0.017}$ &  -- & -- & 4608.79 (4203)\\\hline
\multirow{11}{*}{GX 339-4} & XMM & 654130401 & 1.847$^{+0.035}_{-0.034}$ & 1.416$^{+0.018}_{-0.017}$ &  -- & -- & 3885.28 (1677)\\
 & XMM & 0605610201 & 2.006$^{+0.061}_{-0.06}$ & 0.163$\pm0.005$ &  -- & -- & 2124.54 (1884)\\
 & XMM & 204730301 & 2.607$\pm0.021$ & 0.468$\pm0.005$ &  -- & -- & 3671.23 (1882)\\
 & XMM & 0156760101 & 1.809$\pm0.018$ & 7.284$^{+0.064}_{-0.062}$ &  -- & -- & 5649.49 (1887)\\
 & XMM & 148220201 & 2.466$\pm0.022$ & 1.92$^{+0.021}_{-0.02}$ &  -- & -- & 3151.44 (1886)\\
 & XMM & 692341401 & 2.348$^{+0.063}_{-0.062}$ & 0.252$^{+0.008}_{-0.007}$ &  -- & -- & 2236.82 (1886)\\
 & XMM & 204730201 & 2.538$\pm0.021$ & 0.413$\pm0.004$ &  -- & -- & 3928.23 (1881)\\\cline{2-8}
 & Chandra & 4570 & 1.28$\pm0.034$ & 5.635$^{+0.097}_{-0.095}$ &  -- & -- & 8457.89 (3615)\\
 & Chandra & 4571 & 1.307$\pm0.034$ & 6.13$^{+0.104}_{-0.102}$ &  -- & -- & 7719.27 (3615)\\
 & Chandra & 4420 & 2.464$\pm0.049$ & 2.47$^{+0.057}_{-0.056}$ &  -- & -- & 5806.41 (4179)\\
 & Chandra & 4569 & 1.808$\pm0.019$ & 4.709$^{+0.053}_{-0.052}$ &  -- & -- & 6800.43 (3875)\\\hline
\multirow{4}{*}{Ser X-1} & XMM & 0084020401 & 1.348$^{+0.093}_{-0.092}$ & 1.057$^{+0.05}_{-0.047}$ &  -- & -- & 1964.35 (1881)\\
 & XMM & 0084020501 & 1.66$^{+0.108}_{-0.107}$ & 1.126$^{+0.062}_{-0.058}$ &  -- & -- & 1903.34 (1883)\\
 & XMM & 0084020601 & 1.293$^{+0.061}_{-0.06}$ & 1.007$^{+0.03}_{-0.029}$ &  -- & -- & 2049.14 (1883)\\\cline{2-8}
 & Chandra & 700 & 2.295$\pm0.028$ & 1.464$^{+0.023}_{-0.022}$ &  -- & -- & 5598.15 (4687)\\\hline
\multirow{7}{*}{4U 1636-53} & XMM & 0764180401 & 1.499$\pm0.043$ & 0.102$\pm0.002$ &  -- & -- & 2242.25 (1888)\\
 & XMM & 0500350401 & 1.433$\pm0.019$ & 0.606$\pm0.005$ &  -- & -- & 2880.54 (1886)\\
 & XMM & 0764180301 & 2.17$\pm0.031$ & 0.342$\pm0.005$ &  -- & -- & 2193.04 (1886)\\
 & XMM & 0606070101 & 1.465$\pm0.043$ & 0.607$\pm0.012$ &  -- & -- & 2130.24 (1888)\\
 & XMM & 0764180201 & 1.464$\pm0.019$ & 0.627$\pm0.006$ &  -- & -- & 2819.0 (1883)\\
 & XMM & 0606070301 & 1.59$\pm0.02$ & 0.579$\pm0.005$ &  -- & -- & 2894.44 (1884)\\\cline{2-8}
 & Chandra & 105 & 2.799$\pm0.039$ & 1.345$\pm0.028$ &  -- & -- & 5001.72 (4687)\\\hline
\multirow{1}{*}{IGR J00291+5934} & XMM & 744840201 & 0.468$^{+0.068}_{-0.067}$ & 0.026$\pm0.001$ &  -- & -- & 2171.9 (1881)\\\hline

\label{tab:contfit}
\end{tabular}
\end{table*}
\begin{table*}

\centering

\caption{Fit abundances Information}

\begin{tabular}{l|cc|cccc}
Source Name & Instrument & Obs. ID & $N_H$ (10$^{22}$ cm $^{-2}$)& Ne{\sc i} (10$^{17}$ cm $^{-2}$)& Ne{\sc ii} (10$^{17}$ cm $^{-2}$) & Fe (10$^{17}$ cm $^{-2}$)\\

\hline\hline
\multirow{5}{*}{GX 9+9} & XMM & 0694860301 & 0.241$\pm0.004$ & 2.171$^{+0.237}_{-0.241}$ & 0.759$^{+0.193}_{-0.183}$ & 0.761$\pm0.038$\\
 & XMM & 0090340601 & 0.258$\pm0.004$ & 2.86$^{+0.324}_{-0.331}$ & 0.603$^{+0.252}_{-0.231}$ & 0.731$\pm0.047$\\
 & XMM & 0090340101 & 0.276$\pm0.012$ & 2.672$^{+0.928}_{-0.962}$ & 0.425$^{+0.656}_{-0.425}$ & 0.767$^{+0.154}_{-0.152}$\\\cline{2-7}
 & Chandra & 703 & 0.361$^{+0.026}_{-0.025}$ & 1.519$^{+0.589}_{-0.576}$ & 0.53$^{+0.189}_{-0.173}$ & 0.925$^{+0.135}_{-0.134}$\\
 & Chandra & 11072 & 0.29$^{+0.009}_{-0.008}$ & 1.488$^{+0.254}_{-0.251}$ & 0.434$^{+0.088}_{-0.083}$ & 1.088$\pm0.07$\\\hline
\multirow{4}{*}{Cyg X-2} & XMM & 303280101 & 0.25$\pm0.002$ & 1.63$^{+0.122}_{-0.123}$ & 0.364$^{+0.083}_{-0.08}$ & 0.761$\pm0.021$\\
 & XMM & 111360101 & 0.282$\pm0.003$ & 0.802$^{+0.163}_{-0.165}$ & 0.297$^{+0.114}_{-0.108}$ & 0.857$\pm0.027$\\\cline{2-7}
 & Chandra & 10881 & 0.163$\pm0.006$ & 1.897$^{+0.162}_{-0.161}$ & 0.435$^{+0.059}_{-0.057}$ & 1.084$\pm0.061$\\
 & Chandra & 1102 & 0.301$\pm0.005$ & 1.153$^{+0.358}_{-0.352}$ & 0.719$^{+0.147}_{-0.137}$ & 0.956$^{+0.058}_{-0.057}$\\\hline
\multirow{2}{*}{GX 349+2} & XMM & 0506110101 & 0.955$^{+0.021}_{-0.02}$ & 10.834$^{+0.518}_{-0.525}$ & 2.068$^{+0.43}_{-0.413}$ & 3.397$\pm0.148$\\
 & Chandra & 3354 & 1.318$\pm0.016$ & 9.445$^{+1.065}_{-1.062}$ & 1.472$^{+0.522}_{-0.462}$ & 4.023$^{+0.412}_{-0.411}$\\\hline
\multirow{5}{*}{GRO J1655-40} & XMM & 112921301 & 0.674$\pm0.025$ & 6.45$^{+1.427}_{-1.394}$ & 4.87$^{+1.294}_{-1.27}$ & 2.115$^{+0.301}_{-0.302}$\\
 & XMM & 112921401 & 0.854$^{+0.005}_{-0.004}$ & 9.943$^{+0.258}_{-0.265}$ & 1.236$^{+0.217}_{-0.207}$ & 2.845$\pm0.056$\\
 & XMM & 0155762601 & 0.805$^{+0.013}_{-0.012}$ & 7.918$^{+0.26}_{-0.265}$ & 0.464$^{+0.185}_{-0.171}$ & 2.407$\pm0.075$\\
 & XMM & 112921501 & 0.859$^{+0.005}_{-0.004}$ & 8.965$^{+0.272}_{-0.274}$ & 2.021$^{+0.233}_{-0.229}$ & 2.83$\pm0.056$\\
 & XMM & 0155762501 & 0.701$\pm0.006$ & 8.419$^{+0.264}_{-0.27}$ & 1.235$^{+0.23}_{-0.221}$ & 2.106$\pm0.065$\\\hline
\multirow{1}{*}{XTE J1118+480} & Chandra & 1701 & 0.006$\pm0.003$ & $0.088^*$ & $0.112^*$ & 0.177$\pm0.1$\\\hline
\multirow{5}{*}{4U 1735-44} & XMM & 0090340201 & 0.325$\pm0.004$ & 2.848$^{+0.36}_{-0.367}$ & 0.627$^{+0.281}_{-0.256}$ & 0.901$^{+0.055}_{-0.054}$\\
 & XMM & 0693490201 & 0.302$\pm0.002$ & 3.149$^{+0.165}_{-0.168}$ & 0.652$^{+0.128}_{-0.122}$ & 0.847$\pm0.025$\\\cline{2-7}
 & Chandra & 704 & 0.431$\pm0.009$ & 2.839$^{+0.526}_{-0.516}$ & 0.846$^{+0.242}_{-0.223}$ & 1.265$^{+0.114}_{-0.113}$\\
 & Chandra & 6637 & 0.182$\pm0.011$ & 2.222$^{+0.553}_{-0.547}$ & 0.895$^{+0.251}_{-0.226}$ & 0.807$^{+0.157}_{-0.155}$\\
 & Chandra & 6638 & 0.217$\pm0.012$ & 3.234$^{+0.611}_{-0.599}$ & 0.785$^{+0.234}_{-0.212}$ & 1.332$^{+0.18}_{-0.179}$\\\hline
\multirow{11}{*}{GX 339-4} & XMM & 654130401 & 0.545$\pm0.005$ & 4.105$^{+0.234}_{-0.236}$ & 1.318$^{+0.192}_{-0.187}$ & 1.63$\pm0.044$\\
 & XMM & 0605610201 & 0.57$\pm0.011$ & 3.705$^{+0.735}_{-0.753}$ & 1.304$^{+0.634}_{-0.576}$ & 1.727$\pm0.128$\\
 & XMM & 204730301 & 0.646$\pm0.004$ & 3.363$^{+0.218}_{-0.22}$ & 1.097$^{+0.173}_{-0.167}$ & 1.802$\pm0.042$\\
 & XMM & 0156760101 & 0.603$\pm0.003$ & 5.387$^{+0.235}_{-0.239}$ & 1.411$^{+0.198}_{-0.192}$ & 1.825$\pm0.04$\\
 & XMM & 148220201 & 0.621$\pm0.004$ & 5.431$^{+0.296}_{-0.299}$ & 1.802$^{+0.251}_{-0.244}$ & 1.727$\pm0.046$\\
 & XMM & 692341401 & 0.602$\pm0.012$ & 2.909$^{+0.743}_{-0.739}$ & 2.317$^{+0.629}_{-0.607}$ & 1.574$\pm0.127$\\
 & XMM & 204730201 & 0.633$\pm0.004$ & 3.068$^{+0.23}_{-0.233}$ & 1.269$^{+0.186}_{-0.181}$ & 1.794$\pm0.042$\\\cline{2-7}
 & Chandra & 4570 & 0.511$\pm0.007$ & 3.826$^{+0.185}_{-0.184}$ & 1.32$^{+0.096}_{-0.093}$ & 1.975$^{+0.084}_{-0.083}$\\
 & Chandra & 4571 & 0.521$\pm0.007$ & 3.837$^{+0.18}_{-0.178}$ & 1.095$^{+0.087}_{-0.085}$ & 2.082$\pm0.082$\\
 & Chandra & 4420 & 0.629$\pm0.01$ & 3.563$^{+0.249}_{-0.246}$ & 1.713$^{+0.141}_{-0.136}$ & 1.913$\pm0.081$\\
 & Chandra & 4569 & 0.581$\pm0.005$ & 3.999$^{+0.219}_{-0.217}$ & 1.548$^{+0.122}_{-0.119}$ & 1.761$^{+0.094}_{-0.095}$\\\hline
\multirow{4}{*}{Ser X-1} & XMM & 0084020401 & 0.557$\pm0.018$ & 5.073$^{+1.014}_{-1.106}$ & 0.522$^{+0.886}_{-0.522}$ & 1.43$\pm0.206$\\
 & XMM & 0084020501 & 0.606$\pm0.021$ & 3.886$^{+1.327}_{-1.303}$ & 2.826$^{+1.124}_{-1.085}$ & 1.805$\pm0.236$\\
 & XMM & 0084020601 & 0.537$^{+0.012}_{-0.011}$ & 6.865$^{+0.784}_{-0.873}$ & 0.756$^{+0.751}_{-0.592}$ & 1.73$\pm0.137$\\\cline{2-7}
 & Chandra & 700 & 0.704$\pm0.006$ & 5.382$^{+0.396}_{-0.395}$ & 1.543$^{+0.223}_{-0.214}$ & 2.34$\pm0.098$\\\hline
\multirow{7}{*}{4U 1636-53} & XMM & 0764180401 & 0.317$\pm0.007$ & 4.722$^{+0.585}_{-0.606}$ & 0.691$^{+0.465}_{-0.397}$ & 1.186$\pm0.092$\\
 & XMM & 0500350401 & 0.383$\pm0.003$ & 3.987$^{+0.259}_{-0.263}$ & 0.871$^{+0.202}_{-0.191}$ & 1.144$\pm0.041$\\
 & XMM & 0764180301 & 0.436$\pm0.005$ & 4.372$^{+0.413}_{-0.425}$ & 0.79$^{+0.333}_{-0.302}$ & 1.023$\pm0.063$\\
 & XMM & 0606070101 & 0.384$^{+0.008}_{-0.007}$ & 3.655$^{+0.647}_{-0.673}$ & 1.217$^{+0.587}_{-0.528}$ & 1.226$\pm0.091$\\
 & XMM & 0764180201 & 0.392$\pm0.003$ & 4.275$^{+0.269}_{-0.273}$ & 0.866$^{+0.217}_{-0.205}$ & 1.061$\pm0.041$\\
 & XMM & 0606070301 & 0.406$\pm0.003$ & 4.335$^{+0.271}_{-0.276}$ & 0.997$^{+0.217}_{-0.207}$ & 1.107$\pm0.041$\\\cline{2-7}
 & Chandra & 105 & 0.588$\pm0.009$ & 2.902$^{+0.444}_{-0.436}$ & 0.875$^{+0.21}_{-0.196}$ & 1.831$\pm0.105$\\\hline
\multirow{1}{*}{IGR J00291+5934} & XMM & 744840201 & 0.335$\pm0.013$ & 5.693$^{+0.859}_{-0.91}$ & 0.873$^{+0.738}_{-0.609}$ & 1.589$\pm0.165$\\\hline

\label{tab:abundfit}
\end{tabular}

$^*$ upper limits
\end{table*}

\end{document}